\documentclass{article}
\usepackage{amsmath,graphicx,mlspconf}
\usepackage{multirow}
\usepackage{booktabs}
\usepackage[pagebackref=false]{hyperref}
\usepackage{cleveref}
\usepackage{arydshln}

\usepackage{xcolor}

\expandafter\def\expandafter\normalsize\expandafter{%
    \normalsize%
    \setlength\abovedisplayskip{4pt}%
    \setlength\belowdisplayskip{4pt}%
    \setlength\abovedisplayshortskip{-8pt}%
    \setlength\belowdisplayshortskip{2pt}%
}

\title{View it like a radiologist: shifted windows for deep learning augmentation of CT images}

\name{%
   Eirik A. Østmo$^{1}$%
   \: Kristoffer K. Wickstrøm$^{1}$%
   \: Keyur Radiya$^{1 2}$%
   \: Michael C. Kampffmeyer$^{1 3}$%
   \: Robert Jenssen$^{1 3 4}$\thanks{EØ, KW, MK, RJ are with the UiT Machine Learning group and SFI Visual Intelligence. RJ is also with the Pioneer Centre for AI. This work was supported by The Research Council of Norway [grant no.\ 309439, 315029, 303514].
   We acknowledge Sigma2 – the National Infrastructure for High Performance Computing and Data Storage in Norway, for awarding access to the LUMI supercomputer, owned by the EuroHPC Joint Undertaking.}%
}
\address{%
   $^{1}$ UiT The Arctic University of Norway \quad%
   $^{2}$ University Hospital of North Norway \\%
   $^{3}$ Norwegian Computing Center \quad%
   $^{4}$ University of Copenhagen%
}

\begin{document}

\maketitle

\begin{abstract}
Deep learning has the potential to revolutionize medical practice by automating and performing important tasks like detecting and delineating the size and locations of cancers in medical images. However, most deep learning models rely on augmentation techniques that treat medical images as natural images. For contrast-enhanced Computed Tomography (CT) images in particular, the signals producing the voxel intensities have physical meaning, which is lost during preprocessing and augmentation when treating such images as natural images. To address this, we propose a novel preprocessing and intensity augmentation scheme inspired by how radiologists leverage multiple viewing windows when evaluating CT images. Our proposed method, \textit{window shifting}, randomly places the viewing windows around the region of interest during training. This approach improves liver lesion segmentation performance and robustness on images with poorly timed contrast agent. Our method outperforms classical intensity augmentations as well as the intensity augmentation pipeline of the popular nn-UNet on multiple datasets.

\end{abstract}
\section{Introduction}
\label{sec:intro}

\begin{figure}[t]
  \centering
  \includegraphics[width=1.0\columnwidth]{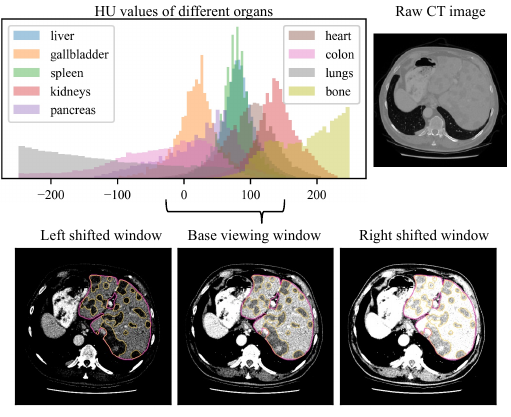}
  \vspace{-0.8cm}
  \caption{The distribution of Hounsfield Units (HU) and visualization of a contrast-enhanced CT image. The liver is delineated (pink), with darker cancer areas (yellow). During preprocessing, a base viewing window is applied to enhance the relevant regions in the scan. By randomly shifting the window during training, we achieve a natural augmentation effect.}
  \label{fig:organ_intensities}
  \vspace{-0.45cm}
\end{figure}

Deep learning (DL) based applications are becoming crucial tools for medical image analysis \cite{wuDeepNeuralNetworks2020}. This includes accurately and robustly segmenting lesions, such as primary and secondary tumors in different organs. For patients with liver cancer, segmentation is crucial for diagnostics, treatment planning, and follow-up \cite{gotraLiverSegmentationIndications2017}. Contrast-enhanced Computed Tomography (CT) imaging of the abdomen is the main source of information, and the images are traditionally assessed manually by radiologists to detect and segment liver lesions.
Although deep learning methods have shown promising results for medical image segmentation, they still face limitations and challenges when applied to liver tumor segmentation \cite{antonelliMedicalSegmentationDecathlon2022, bilicLiverTumorSegmentation2023}. This is especially true when the contrast agent is not timed correctly with the CT image, resulting in poor contrast between healthy liver tissue and the lesion, making it difficult for DL-based methods to perform the task robustly \cite{bilicLiverTumorSegmentation2023}.

Deep learning methods for CT imaging often rely on fixed preprocessing schemes that make CT images appear more like natural images, such that standard methodology can be applied. However, this approach does not respect the physical properties of the CT modality and fails to leverage domain knowledge about how the signal-generating processes in CT imaging, the contrast enhancement, and anatomy interact. This may lead to a loss of information about semantics given by the absolute and relative intensities of different tissues, and naively adapting strategies from other computer vision (CV) domains may thus harm performance.

\begin{figure}[t]
  \centering
  \includegraphics[width=1.0\columnwidth]{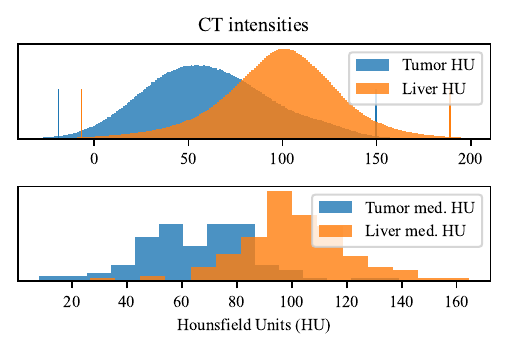}
  \vspace{-0.9cm}
  \caption{\textit{Top}: The global distribution of foreground intensities in the LiTS dataset. The base window and ROI is determined by the lower and upper HU thresholds marked by the vertical bars. \textit{Bottom}: Distribution of per volume median HU for liver and tumor in the LiTS dataset, which is used to determine the range to sample the window levels from during training.}
  \label{fig:lits_intensities}
  \vspace{-0.3cm}
\end{figure}

Inspired by how radiologists apply different viewing settings when examining CT images \cite{sahiValueLiverWindows2014}, we propose a dynamic preprocessing and augmentation scheme for training that enables DL models to learn robustness to poor contrast timing in the CT scan. 
In this paper, we present \textit{window shifting} (\autoref{fig:organ_intensities}), a new method that achieves a CT coherent pixel-wise intensity augmentation by considering the distribution of the region of interest (ROI) (\autoref{fig:lits_intensities}) and dynamically clip the intensities during preprocessing.
We demonstrate the effect of our method in liver lesion segmentation of contrast-enhanced CT images. We show that our clinically motivated augmentation method improves segmentation performance and robustness, and provide analysis and simple heuristics for applying the technique to specific regions.
\section{Background and related work}
\label{sec:background}
\vspace{-0.1cm}

\textbf{Signal-generating process in CT imaging.} \hspace{0.1cm} 
CT is an important tool when screening patients for cancer in the liver. The signal measured in CT imaging is the radiodensity of a volume generated by a rotating X-ray source and captured by a detector. The amount of radiation absorbed throughout the volume is given by the linear attenuation coefficient $\mu$. In medicine, this measurement is converted to Hounsfield units (HU) by a linear transformation that considers the linear attenuation coefficient of water $\mu_{\text{water}}$ and air $\mu_{\text{air}}$, such that the values are scaled and shifted to be meaningful in a medical context. For a given voxel with average linear attenuation, the HU is given by
\begin{equation}
    HU = 1000 \cdot \dfrac{\mu - \mu_{\text{water}}}{\mu_{\text{water}} - \mu_{\text{air}}}
    \label{eq:HU}
\end{equation}
hence, $HU_{\text{air}} = -1000$ and $HU_{\text{water}} = 0$.
To enhance the visibility of certain tissues or structures in the body, contrast enhancement is typically used when acquiring CT scans. In contrast-enhanced CT scans, a radio-opaque contrast agent is injected into the patient's vein and circulates through the body, highlighting specific regions at different times during the CT scan. For example, 50-70 seconds after injection in an upper extremity vein, the contrast agent has reached the portal venous phase, and contrast in the liver is increased. This results in increased measurements of HU in the liver tissue with at least 50 HU \cite{shibamotoInfluenceContrastMaterials2007}, and strong contrast enhancement in the liver images. As many liver lesions respond differently to contrast agents, they become more visible in this phase of the CT image, due to the increased HU values in the healthy liver tissue around them (\autoref{fig:organ_intensities}).

\textbf{CT preprocessing.}\hspace{0.1cm}
To optimize CT images for visual evaluation, the recorded HU in the CT scan are preprocessed by applying a viewing window. The viewing window limits the range of HU by clipping them to a narrower range that covers the ROI. In medical DL applications the viewing window is usually fixed during preprocessing to a suitable range \cite{bilicLiverTumorSegmentation2023, hanAutomaticLiverLesion2017, isenseeNnUNetSelfconfiguringMethod2021} for the ROI. However, in the case of CT liver and tumor segmentation, there is great variation in the clipping ranges used in preprocessing, which suggests that a suboptimal clipping range may be common \cite{bilicLiverTumorSegmentation2023}. To avoid arbitrarily chosen viewing windows, one popular approach is to inspect the foreground intensity distribution of the dataset and set the lower and upper HU boundaries to given percentiles of the distribution \cite{isenseeNnUNetSelfconfiguringMethod2021}. 

Due to differences in timing and tissue response in contrast-enhanced CT images, the liver and tumor intensities vary across different images. Applying a fixed preprocessing strategy to images of varying quality may therefore cause images to look brighter or darker after intensity clipping. A clinician could easily correct this by shifting the viewing window center towards the region of interest. To address this issue in deep learning, models that utilize multiple viewing windows for preprocessing have been proposed for segmentation \cite{kwonTrainableMulticontrastWindowing2020}, as well as self-supervised representation learning \cite{wickstromClinicallyMotivatedSelfsupervised2023}. Choosing a sufficiently large viewing window would also address the issue but may lead to images with lower contrast as many HU values are mapped to similar grayscale levels \cite{leePixelLevelDeepSegmentation2017}. Regardless of the clipping range, after preprocessing CT images are usually treated as non-medical images in common CV applications.

\begin{figure*}[t]
  \centering
  \includegraphics[width=1.0\textwidth]{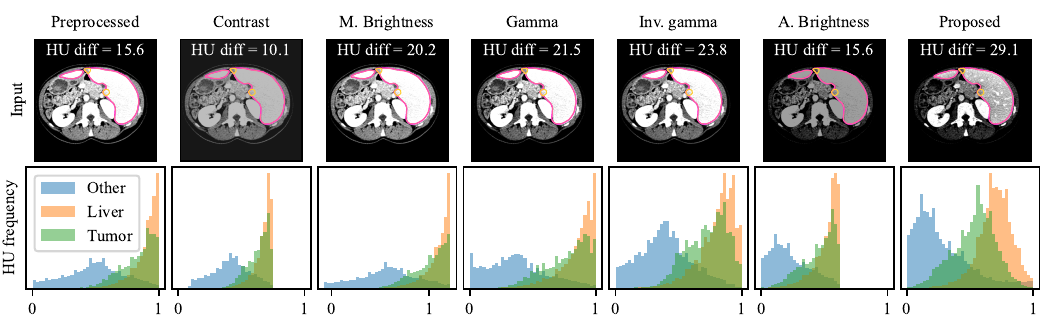}
  \vspace{-1.0cm}
  \caption{Images where the liver responds more than usual to the contrast agent may look overexposed after standard preprocessing. Standard augmentation techniques fail to reintroduce useful variation into such images because much of the original distribution is removed during preprocessing. Our proposed method address this by leveraging more of the original data.}
  \label{fig:aug_intensities}
  \vspace{-0.4cm}
\end{figure*}

\textbf{Intensity augmentation.}\hspace{0.1cm} 
Data augmentations are random transformations that are applied to the model's inputs during training to prevent it from overfitting to the training set by increasing the data variation and diversity \cite{shortenSurveyImageData2019}. Many intensity augmentations developed for RGB images are not applicable for medical images. Color-jittering like changing the hue and saturation requires three input channels (RGB), while CT have only one (grayscale). Many therefore resort to contrast and brightness adjustments as well as gamma correction (\autoref{fig:aug_intensities}), which are linear or non-linear operations that are performed directly on the preprocessed CT intensity values \cite{isenseeNnUNetSelfconfiguringMethod2021, brionDomainAdversarialNetworks2021, hatamizadehUNETRTransformers3D2022}. For natural images, where only the pixels' relative intensities are important, intensity augmentations will teach the model to focus on shapes and textures, rather than absolute intensities, when performing the task. However, in modalities with meaningful pixel intensities the absolute intensities in an image can distinguish between important semantics and should thus be kept intact.

The meaningful signal and the relationship between distinct intensities in the important parts of a CT image may be disrupted by the combination of static preprocessing and classical intensity augmentations (Figure 3). Consequently, a DL model could be prevented from learning representations of shape and texture in combination with meaningful pixel values. A CT specific intensity augmentation has been proposed \cite{kloenneDomainspecificCuesImprove2020} but does not address these issues, and does not provide the variation and smoothness comparable with standard augmentations.

We propose to address the above challenges with a novel preprocessing and augmentation scheme that respects the signal-generating process in CT imaging. Our method dynamically change the preprocessing viewing window during training to achieve a relevant augmentation effect that teaches the model robustness to contrast timing and tissue response.

\vspace{-0.2cm}
\section{Methodology}
\label{sec:methodology}
\vspace{-0.1cm}
In this paper, we propose window shifting, an augmentation scheme that integrates into the preprocessing pipeline of the deep learning model and exposes the model to a continuous range of variations that are relevant for the ROI. We stochastically apply shifted viewing windows around the relevant parts of the intensity distribution, to increase the diversity of training images. This allows us to alleviate standard pixel distortions such as brightness and contrast augmentations completely during training, and learn more robust and clinically relevant features, that further improves segmentation results.

Window shifting is intuitively simple and can be tailored to the given task hand by analyzing the data at hand. The implementation is efficient and reduces the number of operations needed. This is possible due to the joint preprocessing and augmentation step. 
\let\thefootnote\relax\footnotetext{Code at \url{https://github.com/agnalt/window-shifting}.}
\vspace{-0.3cm}

\subsection{Window shifting for CT intensity augmentations}
\label{sec:window_shifting}
Window shifting use a base viewing window that covers 99 \% of the dataset's foreground distribution. The lower and upper bounds of the base window are determined by the 0.5 and 99.5 percentile of the foreground intensities \cite{isenseeNnUNetSelfconfiguringMethod2021}. We define the viewing window that covers the clipping range by a window width $W$ and a centered window level $L$. During preprocessing the HU intensities are clipped such that
\begin{align}
    HU &\in [L - \frac{W}{2}, L + \frac{W}{2}].
    \label{eq:preprocess}
\end{align}
For illustration, \autoref{fig:lits_intensities} displays the distribution of foreground tumor and liver pixels, as well as the boundaries, in the Liver tumor segmentation (LiTS) dataset \cite{bilicLiverTumorSegmentation2023}.

During training, pixel-wise preprocessing of the HU intensities is postponed until augmentation. To achieve the augmentation effect, a substitute window level is sampled with a probability $p$ from a uniform distribution
\begin{equation}
    L \sim \text{Uniform}(L_{\text{low}}, L_{\text{high}}), \tag{3}
\end{equation}
where $L_{\text{low}}$ and $L_{\text{high}}$ are the lower and upper bound. $L_{\text{low}}$ and $L_{\text{high}}$ are defined from the distribution of median foreground pixels per case in the dataset, and $L_{\text{low}} < L < L_{\text{high}}$
As auxiliary viewing windows used during augmentation are sampled around the level $L$, and we sample windows with $p < 1$, the most common window the model is exposed to will be the base window. This enables us to simplify preprocessing during model inference, and resort to the base viewing window.
\begin{table*}[t]
\centering
\small
\scalebox{1.0}{
\begin{tabular}{rrrrr}
\hline
Intensity augmentation & LiTS tumor dice        & Hepatic tumor dice     & UNN tumor dice         & Mean tumor dice        \\ \hline
None (geometric only)  & 0.574 ± 0.202          & 0.402 ± 0.127          & 0.366 ± 0.133          & 0.447 ± 0.154          \\
Mult. brightness       & 0.639 ± 0.087          & 0.419 ± 0.036          & 0.452 ± 0.048          & 0.503 ± 0.057          \\
Contrast               & 0.634 ± 0.086          & 0.445 ± 0.044          & 0.451 ± 0.072          & 0.510 ± 0.067          \\
Gamma                  & 0.630 ± 0.078          & 0.472 ± 0.041          & 0.459 ± 0.050          & 0.520 ± 0.056          \\
Add. brightness        & 0.636 ± 0.106          & 0.474 ± 0.030          & 0.519 ± 0.050          & 0.543 ± 0.062          \\
Gamma inverse          & \textbf{0.650 ± 0.089} & 0.494 ± 0.030          & 0.521 ± 0.048          & 0.555 ± 0.056          \\ \hdashline
nn-UNet (baseline)       & 0.638 ± 0.088          & 0.483 ± 0.035          & 0.522 ± 0.052          & 0.548 ± 0.058          \\ \hdashline
Window shifting (proposed)       & \textbf{0.650 ± 0.086} & \textbf{0.513 ± 0.030} & \textbf{0.532 ± 0.026} & \textbf{0.565 ± 0.047} \\ \hline
\end{tabular}}
\vspace{-0.1cm}
\caption{The effect of individual intensity augmentations measured by the dice coefficient on the validation split of the LiTS dataset, as well as the transfer performance of tumor segmentation to the Hepatic vessels and the UNN dataset.}
\label{tab:individual_augmentations}
\vspace{-0.4cm}
\end{table*}
\vspace{-0.6cm}
\subsection{Determining window shifting boundaries}
\label{sec:window_bounds}
As the global intensity distribution is used to determine the base window, standard preprocessing might be biased to favour good visualization of cases with large tumors. To counteract this, we use the per-volume median HU to determine how much it is beneficial to shift the window level in augmentation. \autoref{fig:lits_intensities} illustrates how much the HU values for tumor and liver tissue vary between different CT scans in the LiTS dataset, and thus how much the window should be adjusted during training. Based on the per-volume median intensity distribution we set the lower and upper bounds of the range $[L_{\text{low}}, L_{\text{high}}]$ that are used during augmentation to the 0.5 and 99.5 percentile.

\subsection{Effect of window shifting}
When applying the viewing window during preprocessing, much of the image's original intensity distribution is removed. For most preprocessed CT images, this results in better contrast in the relevant regions of the intensity distribution, which improves the model's capability to identify important nuances for the segmentation task.
However, for images where the tissue responds more or less than usual, or if the timing is off, the preprocessed images might come out looking over- or underexposed (\autoref{fig:aug_intensities}). In such cases, the model may have trouble extracting useful information and learn from the given image. Traditional intensity augmentations applied during training often have trouble with introducing sufficient and relevant variations to such corrupted images, as they are applied on the preprocessed image. \autoref{fig:aug_intensities} displays how standard augmentation schemes alter the intensity distribution, but fails to introduce relevant variations and separate the tumor and liver distribution. We measure the mean HU difference between liver and tumor tissue to evaluate if relevant contrast is introduced. 
The improved visualized results of the proposed method is due to its ability to leverage more of the intensity distribution, that would otherwise be removed during preprocessing.

\section{Experiments and results}
\label{sec:experiments}
We hypothesize that our proposed preprocessing and augmentation scheme lead to models with better generalization performance to cases that are difficult. We evaluate our window shifting method against common individual intensity augmentations, as well as the strong nn-UNet intensity augmentation baseline \cite{isenseeNnUNetSelfconfiguringMethod2021} by training a DL segmentation model to segment liver lesions in contrast-enhanced CT images.

\subsection{Experimental setup}
As the baseline augmentation techniques require preprocessed images, we apply the base viewing window and clip the HU intensities to the 0.5 and 99.5 percentiles of foreground pixels of the dataset. The preprocessed images $\mathbf{x}$ are then scaled and shifted such that $x \in [0, 1]$. In all experiments we apply basic geometric augmentations, like flipping and crop-and-resize, as they are crucial for good performance.

When we compare with baseline intensity augmentations like additive brightness, $\mathbf{x_{\text{new}}} = \mathbf{x} + \alpha$, gamma, $\mathbf{x_{\text{new}}} = \mathbf{x}^{\gamma}$, and inverse gamma correction, $(1- \mathbf{x_{\text{new}}}) = (1-\mathbf{x})^{\gamma}$, they are applied before the images are z-score normalized using the global dataset foreground mean and standard deviations. Multiplicative brightness and contrast augmentation, where the pixel intensities are scaled $\mathbf{x_{\text{new}}} = \mathbf{x} \cdot \beta$, require normalized values and are thus applied after normalization. For contrast augmentation, the scaled pixel values are clipped to their original range after scaling. Window shifting is applied as described in \autoref{sec:window_shifting} and \ref{sec:window_bounds} before the HU units are scaled to the range $[0, 1]$ and normalized.

To perform the segmentation task, we use the DeepLabv3+ segmentation model with a Resnet50 backbone \cite{chenEncoderDecoderAtrousSeparable2018}. This model is similar to the U-net \cite{ronnebergerUNetConvolutionalNetworks2015}, popular in medical imaging, but uses more aggressive upsampling between skip connections, has multiple dilated convolutions to capture multi-scale information, and achieve similar performance to the U-net on multiple datasets \cite{dacruzKidneyTumorSegmentation2022, hadinataCrackDetectionConcrete2021}.

The training and validation data are from the LiTS \cite{bilicLiverTumorSegmentation2023} benchmark, which was also used in the Medical Segmentation Decathlon (MSD) \cite{antonelliMedicalSegmentationDecathlon2022}. The dataset consists of 131 contrast-enhanced CT images of the abdomen gathered from IRCAD Hôpitaux Universitaires, Strasbourg, France. Most images have liver cancers that are either primary or secondary liver tumors. For testing we leverage the Hepatic Vessels dataset from MSD \cite{antonelliMedicalSegmentationDecathlon2022}, which is a completely disjoint dataset with 303 contrast-enhanced abdomen CT images from Memorial Sloan Kettering Cancer Center, New York, US. The tumors in this dataset are also both primary and secondary liver cancers. Complementary to these datasets, we test the transfer performance on the UNN liver tumor dataset, which is under development as a collaboration between UiT The Arctic Univserity of Norway and the University Hospital of North-Norway (UNN). It consists of 27 contrast-enhanced CT volumes of the abdomen from different patients with liver metastasis.

We perform 3 times repeated 4-fold cross-validation on the LiTS dataset, where each split consists of 98/99 training volumes and 33/32 validation volumes. As the aim is to test the effect of different augmentation schemes all models are trained from randomly initialized weights, i.e. without any pretraining, to not bias the results towards pretraining augmentations. We measure the overlap between the predicted segmentation mask and the ground truth label using the dice coefficient, given by equation
\begin{equation}
    \text{Dice coefficient} = \dfrac{2TP}{2TP + FP + FN}
    \label{eq:dice_coefficient}
\end{equation}
where $TP$, $FP$ and $FN$ are the true positive, false positive, and false negative pixel predictions respectively. We report the mean over all 12 folds as well as the standard deviation.

The segmentation network takes as input the $256 \times 256$ downsized single channel axial slices of the 3D volumes in the training and validation split. For simplicity, and because we only aim to segment the lesion class, only axial slices containing the liver are used. We train with a batch size of 128 for 100 training steps $\times$ 50 epochs. We use an initial learning rate of 0.01, Adam optimizer, 5 warm-up epochs and cosine decay of the learning rate. For all experiments, the geometric augmentations are identical, that is crop-and-resize with probability $p=0.2$ and flipping with $p=0.5$. All intensity augmentations, including window shifting are applied with probability $p=0.3$. The nn-UNet intensity augmentation baseline consists of multiplicative brightness, contrast, gamma correction and inverse gamma applied each with $p=0.15$ \cite{isenseeNnUNetSelfconfiguringMethod2021}. We adopt the strengths of the strong nn-UNet baseline for the augmentations present in its augmentation pipeline. For comparable results, additive brightness is applied with an equivalent distribution shift as our proposed method.

\subsection{Intensity augmentations in CT images}
Many DL algorithms that use augmentations apply multiple intensity augmentations in sequence. To understand and test the relevance and effect of each intensity augmentation, we isolate the individual augmentations and evaluate the results after training with each one. We compare and evaluate the performance with two common scaling augmentations, multiplicative brightness and contrast correction, and two non-linear augmentations, gamma and inverse gamma transformation. In addition, we evaluate the closest standard alternative to window shifting, namely additive brightness, which shifts the distribution of preprocessed pixel values similarly as our method. As a baseline we also evaluate the performance of not applying any intensity augmentations.

We present the results in \autoref{tab:individual_augmentations}, and verify that any intensity augmentation improve segmentation results over the baseline of not applying any.
The proposed method of applying shifted viewing windows during preprocessing increase the segmentation performance more than any of the alternatives. Non-linear gamma transformations, have a more positive effect on segmentation results than the scaling augmentations. This might be due to their ability to stretch out certain parts of the distribution and increase contrast in the more relevant areas yielding more diverse and informative training images, as seen in \autoref{fig:aug_intensities}. 

\subsection{Identifying difficult cases of tumor segmentation}
\label{sec:difficult_tumors}
Small liver tumors and images with low contrast between tumor and healthy liver tissue are identified as especially difficult for liver lesion segmentation models \cite{bilicLiverTumorSegmentation2023}. Specifically, images where the absolute HU difference between healthy and tumor tissue is below 20 HU are considered hard to segment. As low contrast within the liver often is due to poor timing in the contrast-enhanced CT image, we hypothesize that our clinically motivated window shifting augmentation will allow the model to generalize better to such difficult cases.

We identify the difficult volumes in each validation split, where the difference between mean HU intensity in the liver and tumor tissue is $ < 20 $ HU. We consider these images as difficult cases and evaluate the trained model from each fold on these images, and similarly for the UNN dataset, and evaluate the models from all folds on these images to test if this effect generalizes to another dataset.
\begin{table}[t]
\centering
\small
\scalebox{1.0}{
\begin{tabular}{rcc}
\hline
Intensity augmentation     & \multicolumn{1}{r}{LiTS difficult} & \multicolumn{1}{r}{UNN difficult} \\ \hline
None (geometric only)      & 0.231 ± 0.203                             & 0.219 ± 0.148                            \\
Contrast                   & 0.280 ± 0.168                             & 0.355 ± 0.113                            \\
Mult. brightness           & 0.271 ± 0.184                             & 0.363 ± 0.087                            \\
Gamma                      & 0.229 ± 0.162                             & 0.383 ± 0.075                            \\
Gamma inverse              & 0.248 ± 0.155                             & 0.478 ± 0.065                            \\
Add. brightness            & 0.293 ± 0.184                             & 0.485 ± 0.080                            \\ \hdashline
nn-UNet                    & 0.272 ± 0.188                             & 0.501 ± 0.059                            \\ \hdashline
Window shifting            & \textbf{0.303 ± 0.188}                    & \textbf{0.520 ± 0.027}                   \\ \hline
\end{tabular}}
\vspace{-0.1cm}
\caption{The segmentation dice scores on difficult volumes with low contrast ($<20$ HU) between tumor and healthy tissue in the LiTS validation set and the UNN dataset.}
\label{tab:difficult_tumors}
\vspace{-0.4cm}
\end{table}

We report the results of this experiment in \autoref{tab:difficult_tumors}. As the model is tested on only the most difficult cases in each dataset, the overall performance drops as expected. However, the performance of our proposed window shifting is better than its baseline alternatives, which strengthens the main hypothesis. \autoref{tab:difficult_tumors} also suggests that intensity shifting in general is beneficial to simulate the difficult images, as shifting, performed also in additive brightness, does better than in \autoref{tab:individual_augmentations}. 

\section{Conclusion and future work}
\label{sec:conclusion}
In this paper we have proposed a new combined preprocessing and intensity augmentation technique, called window shifting, for DL-based segmentation of contrast-enhanced CT images. Our method respects the physical properties of CT images and improves model segmentation performance and robustness to images with poorly timed contrast agent. We have demonstrated the effectiveness of our proposed method and compared it with classical intensity augmentations as well as nn-UNet’s augmentation pipeline. Our method outperforms them and yields better transfer performance when tested on multiple other CT liver tumor datasets. This clinically motivated augmentation method integrates into the preprocessing pipeline of the DL training framework. Also, we provided an analysis on the choice of hyperparameters for region-specific augmentations. Overall, our proposed window shifting technique is a powerful approach to improve the accuracy and robustness of DL-based segmentation of liver tumors in contrast-enhaced images. We envision that further development of preprocessing and augmentation schemes for DL that respect the underlying signal-generating process may be fruitful arenas for future research with respect to medical images beyond CT, or for images from vastly different areas such as e.g. remote sensing.

\bibliographystyle{IEEEbib}

\bibliography{main}

\end{document}